# Scalpel: The Python Static Analysis Framework

Li Li, Jiawei Wang, Haowei Quan

Monash University, Australia
{`li.li, jiawei.wang1, haowei.quan`}@monash.edu

**Abstract.** Despite being the most popular programming language, Python has not yet received enough attention from the community. To the best of our knowledge, there is no general static analysis framework proposed to facilitate the implementation of dedicated Python static analyzers. To fill this gap, we design and implement such a framework (named Scalpel) and make it publicly available as an open-source project. The Scalpel framework has already integrated a number of fundamental static analysis functions (e.g., call graph constructions, control-flow graph constructions, alias analysis, etc.) that are ready to be reused by developers to implement client applications focusing on statically resolving dedicated Python problems such as detecting bugs or fixing vulnerabilities.

**Call for Contributions:** The Scalpel Python Static Analysis Framework is available at GitHub[1] and is currently under active development. Your contributions to this project, e.g., through reporting bugs, submitting pull requests for fixing bugs or adding new features, improving API documentation and the user guide, will be very much welcomed and appreciated.

## 1 Introduction

Python, invented by Dutch programmer Guido van Rossum back in the late 1980s, is designed to be an interpreted, object-oriented, high-level, general-purpose programming language with design philosophies emphasizing features such as execution efficiency, code readability, etc. It has now become one of the most popular programming languages in the past few years. Recently, it even takes first place from C and became the most popular programming language, based on the latest TIOBE index (Feburary 2022). TIOBE[2] is a software testing company that releases, and update monthly, the so-called TIOBE index about programming language popularity rankings based on searches for those languages on popular websites and search engines.

As Python becomes more and more popular, the number of Python projects is also fast-increasing. Unfortunately, the rapidly increasing momentum has not yet attracted adequate attention from the research community in which there are not many advanced approaches proposed to help developers develop secure and high-quality Python programs. For example, Wang et al. [1] have experimentally found that API deprecation has been poorly handled by Python library contributors through empirically looking into six reputed Python libraries and 1,200 Github Python projects that have accessed into these six libraries. Similarly, Rak-amnouykit et al. [2] have also experimentally revealed that popular Python type checking tools frequently disagree with each other when applied to analyze 173 thousand Python files collected from GitHub. In addition, our community also reports empirical findings on dependency conflicts among Python open-source software [3].

---
[1] https://github.com/SMAT-Lab/Scalpel
[2] https://www.tiobe.com/tiobe-index/

The lack of appropriate static analyzing tools has also been reflected by the fact that the quality of any Python code in publicly released projects is very poor. As experimentally discovered by Wang et al. [4], many of their studied Python code snippets extracted from publicly released Jupyter notebooks contain poor quality code, such as having unused variables or accessing deprecated functions. The authors have further experimentally disclosed that the state-of-the-art Python-based Jupyter notebooks have many other issues that hinder the reproducibility of the notebooks that are designed to be reproducible [5,6]. The same results have also been experimentally confirmed by Pimentel et al. [7], who has reported that the reproducibility rate of Python-based Jupyter notebooks is quite low, i.e., only less than 25% of notebooks can be successfully reproduced.

## 2   Scalpel Overview

We aim to provide Scalpel as a generic Python static analysis framework that includes as many functions as possible (e.g., to easily build inter-function control-flow graph, to interpret the import relationship of different Python modules, etc.) towards facilitating developers to implement their dedicated problem-focused static analyzers.

Fig. 1 depicts the current architecture design of the Scalpel framework, which is three-layered. The bottom layer represents the Python execution environment, which is the only requirement to run Scalpel since it is written in Python. The middle layer represents the framework itself, which contains not only core static analysis functions but also their corresponding test cases accompanied with usage examples. The top layer presents client applications (for statically addressing dedicated Python code issues such as detecting bugs or identifying vulnerabilities) that could be quickly implemented based on the Scalpel framework.

We now briefly introduce the core functions included in the Scalpel framework.

### 2.1   Code Rewriter.

The code rewriter module is designed as a fundamental function for supporting systematic changes of existing Python programs. Two preliminary usages of this function are to (1) simplify the programs for better static analysis and (2) optimize or repair problematic programs.

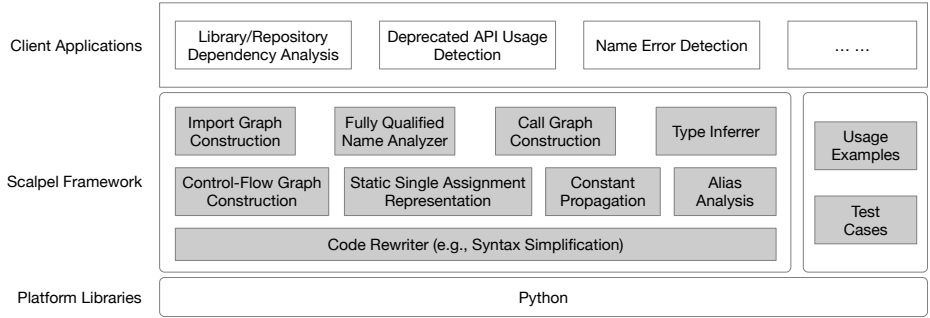

Fig. 1: Overview of the Scalpel Python Static Analysis Framework.

For supporting the first usage, we integrate into the framework a database including a set of rules indicating how matched code snippets should be transformed. As shown in Fig. 2, this database should be continuously extended to fulfill the complicated simplification requirements for achieving effective static analysis of Python programs. At the moment, the Scalpel framework offers five rules for simplifying code syntax (as summarized in Table 1) so as to ease the implementation of subsequent static analysis approaches. For example, if Lambda statements are explicitly transformed to function definitions (e.g., transforming *x = lambda a: a + 10* to *def x(a): return a + 10;*), the corresponding static analysis approaches do not need to worry about the impact of lambda statements, which otherwise will be regarded as assignment statements and hence treat the function references as local variables. Such mistakes will, unfortunately, lead to incorrect results and hence impact the performance of the static analysis approaches.

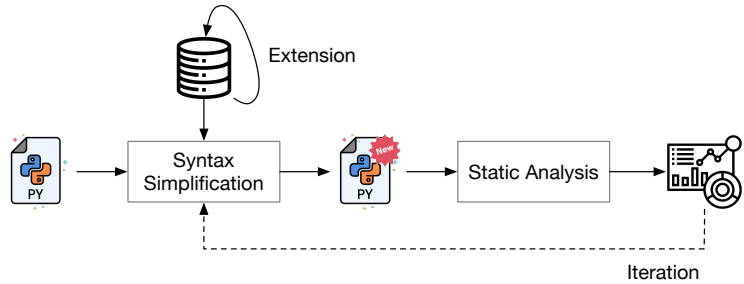

Fig. 2: Code Rewrite by Syntax Simplification..

Another benefit of this syntax simplification approach is that it can be called iteratively if the static analysis results can be leveraged to further simplify the code syntax. Taking *rule 5* as an example, given statement *f1().f2().f3()*, ideally, this rule would like to transform it to three function call statements (i.e., x = f1(); y = x.f2(); z = y.f3()). However, the return type of functions *f1()* and *f2()* may not be known at the beginning of the syntax simplification step, resulting in a non-simplified statement and hence bringing no benefits to the subsequent static analysis approach. However, if the static analysis approach produces information about the return type of functions *f1()* and *f2()*, this information could then help in separating the chained method calls. Subsequently, the separated (simplified) statements may further improve the performance of the static analysis approach. This iteration process can therefore be continuously applied until no further improvements can be observed in the results of the static analysis approach.

For supporting the second usage, inspired by the optimization mechanism provided by Soot (one of the most famous static Java program analysis frameworks [8,9,10]), we also set up a transformation process with dedicated callback methods to be rewritten by users to optimize Python code based on their customized needs. We expect this function will be higly useful to support automated Python code repairs, for both security issues and quality defects.

Table 1: Code Syntax Simplification Rules.

| No | Rule | Before | After |
|---|---|---|---|
| 1 | Container Comprehension Unfolding | lst = [i for i in range(N)] | lst = [] <br> for i in range(N): <br>     lst.append(i) |
| 2 | Nested Function Call handling | x = funA(funB()) | _ret = funB() <br> x = funA(_ret) |
| 3 | Subscription Assignment | x = funA()[1:10] | _ret = funA() <br> x = _ret[1:10] |
| 4 | Lambda Expression Conversion | fun = lambda x: x+1 | def fun <br>     return x+1 |
| 5 | Call Chain Splitting | f1().f2().f3() | x=f1() <br> y=x.f2() <br> z=y.f3() |

### 2.2  Static Analysis Functions

- **Function 1: Control-Flow Graph (CFG) Construction.** The control-flow graph(CFG) construction module generates intra-procedural CFGs, which are an essential component in static flow analysis with applications such as program optimization and taint analysis. A CFG represents all paths that might be traversed through a program during its execution. The CFGs of a Python project can be combined with the call graph to generate an inter-procedural CFG of the project.
  More details can be found in Appendix A.

- **Function 2: Static Single Assignment (SSA) Representation.** The static single assignment module provides compiler-level intermediate representations (IR) for code analysis. By renaming each variable assignment with different names, we are able to obtain explicit use-def chains, therefore precisely tracking how data flow in the program. It can not only be used for dead code elimination, value numbering, but also for constant propagation.
  More details can be found in Appendix B.
- **Function 3: Constant Propagation.** The constant propagation module will evaluate the actual values for variables at certain program points in different execution paths before runtime. With the actual values known beforehand, we are able to optimize code and detect bugs. The constant propagation will utilize the representation from the SSA module to keep recording values from each assignment for a single variable.
  More details can be found in Appendix B.
- **Function 4: Alias Analysis.** Since variables can point to the same memory location or identical values, the alias analysis function is designed to model such usages. This function can be vital to sound constant propagation. In addition, alias analysis will also benefit type checking as well as API name qualifying.
  More details can be found in Appendix B.
- **Function 5: Import Graph Construction.** In Python language, import flows and relations have been pointed out to be important for API mapping[6], dependency analysis. Our Import graph construction aims to provide a data structure to represent these import relationships across the Python module files in the same project. The import graphs of multiple Python projects can be combined to perform inter-library dataflow analysis.
  More details can be found in Appendix C.
- **Function 6: Fully Qualified Name Inferrer.** Python APIs or function names can be invoked in different ways depending on how they are imported. However, this results in inconveniences to API usage analysis. In this module, we will convert all function call names to their fully-qualified names, which are dotted strings that can represent the path from the top-level module down to the object itself. Various tasks can be benefited from this functionality, such as understanding deprecated API usage, dependency parsing as well as building sound call graphs.
  More details can be found in Appendix C.
- **Function 7: Call Graph (CG) Construction.** A call graph depicts calling relationships between methods in a software program. It is a fundamental component in static flow analysis and can be leveraged in tasks such as profiling, vulnerability propagation, and refactoring. This module addresses the challenges brought by complicated features adopted in Python, such as higher-order functions and nested function definitions, to construct the precise call graphs for given Python projects.
  More details can be found in Appendix D.
- **Function 8: Type Inference.** Python, as a dynamically typed language, faces the problem of being hard to utilize the full power of traditional static analysis. This module infers the type information of all variables, including function return values and function parameters in a Python program, making more sophisticated static analysis possible for Python. We utilize backward data-flow analysis and a set of heuristic rules to achieve high precision. More details can be found in Appendix E.

# Appendix

## A  CFG Construction

### A.1  Introduction

*scalpel.cfg* module is used to construct the control flow graphs for given python programs. The basic unit in the CFG, Block, contains a list of sequential statements that can be executed in a program without any control jumps. The 'Block's are linked by 'Link' objects, which represent control flow jumps between two blocks and contain the jump conditions in the form of an expression. The two components are the fundamental data structures in the control flow graph module (*scalpel.cfg*).

### A.2  How to use Control Flow Graph

We will use the following python program to show the CFG construction with *Scalpel*. The piece of code generates the Fibonacci sequence.

```python
# example.py
def fib():
    a, b = 0, 1
    while True:
        yield a
        a, b = b, a + b

fib_gen = fib()
for _ in range(10):
    next(fib_gen)
```

To build the CFG of a source code string/source file, simply import *CFGBuilder* in *scalpel.cfg* and use *build_from_srcname*, $file\_path$. Other functions can also be used to build a CFG from a Python AST tree or a source file.

```python
from scalpel.cfg import CFGBuilder

cfg = CFGBuilder().build_from_file('example.py', './example.py')
```

This returns the CFG for the code in *example.py* in the *cfg* variable. The built CFG can be visualized with *build_visual*:

```python
cfg.build_visual('pdf')
```

Fig. 3 is the produced *exampleCFG.pdf*.

Apart from generating the visual graph, the CFG can be used for many other static analysis purpose. For example, *get_calls* can be used to get all function calls in each block.

### A.3  Visualizations of CFG objects

Scalpel offers some functionalities to render the CFG diagrams into PDF, PNG, JPG or SVG. Please notice this requires the graphviz package installed on your

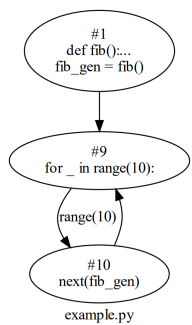

Fig. 3: The control flow graph for the given source files

```
1   for block in cfg:
2       calls = block.get_calls()
```

computer. Please refer to https://graphviz.readthedocs.io/en/stable/manual.html to get it installed.

Developers can use *cfg.build_visual()* to build a *graphviz.dot.Digraph* object, then call *render* function to generate the image file as shown in the following code snippet.

```
1   dot = fun_cfg.build_visual('png')
2   dot.render("cfg_diagram", view=False)
```

### A.4  Visiting Function CFGs

In theory, there is no control flow constraint between sub-procedures such as functions. Therefore, Scalpel generates control flow graphs for every function in the given source files. Considering the nested structure of Python class and function definition, we integrate a recursive data structure for storing control flow graphs.

Take the above source code for example, scalpel will generate two CFGs. One is shown in Fig. 3, and the other is for the function *fib*.

We can use the following way to visit all the function CFGs in the given source. Please note that function CFGs can be indexed by both function name and its block id. This is due to the fact that Python allows users to define functions with same names in the same domain such as:

The implementation of the function *solve* can be different depending on the actual condition. Therefore, we need more than the function name to index a

```
1   x = 0
2   if some condition:
3       def solve():
4           ...
5   else:
6       def solve():
7           ...
8   solve(x)
```

function CFG. Now, if we take the Fig. 3 for example, we can visit all function CFGs in the following way and try to create a PDF file for the diagram of the CFG of function *fib*:

```
1       for (block_id, fun_name), fun_cfg in cfg.functioncfgs.items():
2           ... # do something
3           if fun_name == "fib":
4               graph = fun_cfg.build_visual('png')
5               graph.render("fig_cfg", view=False)
```

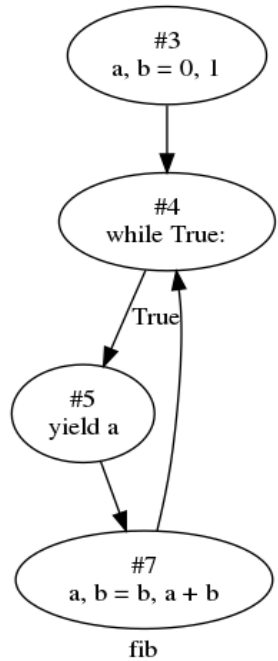

Fig. 4: The control flow graph for the function *fib*

Please note, as there might be functions defined inside a function definition, you can continue perform the similar operation on the *fun_cfg* to retrieve the nested function CFGs. These CFGs can be combined for analysis.

The tutorial code can be found at: `https://github.com/SMAT-Lab/Scalpel/tree/master/examples/cfg_tutorial.py`

### A.5  APIs

The API documentation can be found at `https://smat-lab.github.io/Scalpel/scalpel/cfg.html`

## B  SSA Representations, Constant Propagation and Alias Analysis

### B.1  Introduction

Static Single Assignment (SSA) is a technique of IR in the compiling theory[11], it also shows great benefits to static program analysis tasks such as constant propagation, dead code elimination by providing explicit def-use chains while reducing the computational complexity .

Constant propagation is also a matured technique in static analysis. It is the process of evaluating or recognizing the actual constant values or expressions at a particular program point. This is realized by utilizing control flow and data flow information. Determining the possible values for variables before runtime greatly benefits software analysis. For instance, with conditional constant value propagation, we can detect and remove dead code or perform type checking.

In Scalpel, we implement constant propagation along with the SSA for execution efficiency.

### B.2  How to use the SSA and Constant Propagation module

The demo input python program we will be using is as follows.

```
1  code_str="""
2  c = 10
3  a = −1
4  if c>0:
5      a = a+1
6  else:
7      a = 0
8  total = c+a
9  """
```

It can be seen from the above code, the variable *a* has two possible values. By utilizing the phi functions in SSA, we are able to infer that the actual return value will have two potential values. The input parameter for SSA computing is the CFG as it represents how the code blocks are organized in the program execution flow.

Please note that the function *compute_SSA* returns two dictionaries. For the first one, the key values are block numbers in the given CFG, the value is a list of dictionaries, each of which corresponds to one statement in the block. For instance, the *ssa_results[3]* (block id is 3 in this case) is a list of SSA representations for the last block. If we inspect the last block (the variable *total* is assigned), the results are

```
1    3: [{'c': {0}, 'a': {1, 2}}]
```

This is due to that the variable *a* can take values from two assignments. This can be easily observed from the following diagram.

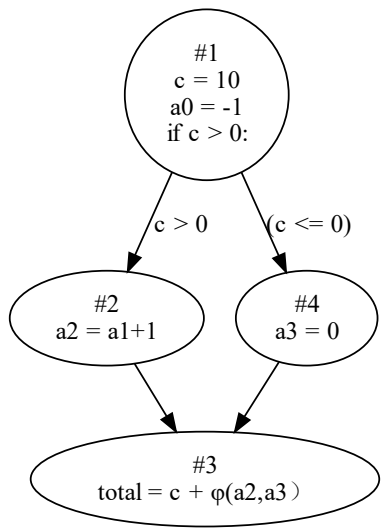

Fig. 5: An example of static single assignment representations in which the variable *a* in the block #3 takes two possible values from the incoming block #2 and #4.

The second one named *const_dict* is the global constant values for the numbered identifiers. For instance, *const_dict[(a,0)]* is the constant value after the first assignment to variable *a*. The constant values in this module are instances of Python *ast.expr*. In this particular case, *(a,0)* is an *ast.BinOp* type and *(a,1)* is an *ast.Num* type.

The tutorial code can be found at: `https://github.com/SMAT-Lab/Scalpel/blob/master/examples/ssa_example.py`

### B.3   Alias Analysis

As stated above, the *const_dict* stores all the values that are assigned to given variables (renamed in this situation), therefore, alias pairs can be implemented in a simple way. The idea is to scan all the values for a particular variable and see if the value assigned to the variable is an *ast.Name* object. The following code snippet shows how to find out the alias pair.

```
1    ssa_results, const_dict = m_ssa.compute_SSA(cfg)
2    alias_name_pairs = []
3    for name, value in const_dict.items():
4        if isinstance(value, ast.Name):
5            alias_name_pairs.append((name, value.id))
```

### B.4   APIs

The API documentation can be found at `https://smat-lab.github.io/Scalpel/scalpel/SSA.html`

## C   Import Graph Construction

### C.1   Introduction

*scalpel.import_graph* creates a data structure for describing import relationships of a python project. An import graph represents the dependency relationship of module files in the given project. This information can be important for understanding the import flow, hierarchy, encapsulation as well as software architecture. Each node in the import graph data structure is a module file that can be manipulated to extract statements and function calls. All the leaf nodes in the import graph can be further processed.

### C.2   How to use Import Graph

Given three example python modules in the following example folder where three Python module files are defined.

```
1    |-- example
2        |-- module_a.py
3        |-- module_b.py
4        |-- module_c.py
```

```
1    #/example/module_b.py
2    from .module_b import B
3    from .module_c import C
4
5    class A:
6        def foo(self):
7            return
```

```
1    #/example/module_c.py
2    from .module_c import C
3
4    class B:
5        def foo(self):
6            return
```

```
1    import os
2
3    class C:
4        def foo(self):
5            return
```

To build the import graph of the package, import and use *Tree* and *ImportGraph* in *scalpel.import_graph.import_graph*.

```
1    from scalpel.import_graph.import_graph import Tree,ImportGraph
2
3    root_node = Tree("import_graph_example_pkg")
4    import_graph = ImportGraph()
5    import_graph.build_dir_tree()
6    module_dict = import_graph.parse_import(root_node)
7    leaf_nodes = import_graph.get_leaf_nodes()
8    print(len(leaf_nodes))
```

For each of leaf notes, we can future to extract its type information, function definition list or more meta information. The tutorial code can be found at:
`https://github.com/SMAT-Lab/Scalpel/blob/master/examples/import_graph_tutorial.py`

### C.3 APIs

The API documentation can be found at `https://smat-lab.github.io/Scalpel/scalpel/import_graph.html`

## D Call Graph Construction

### D.1 Introduction

A call graph depicts calling relationships between subroutines in a computer program. It is an essential component in most static analysis and can be leveraged to build more sophisticated applications such as profiling, venerability propagation and refactoring.

Please note, *scalpel.call_graph.pycg* module is a wrapper of PyCG[12] at present. It aims to construct the call graphs for given Python projects. The basic node can be either a function, a class, or a module. The edges represent calling relationships between program nodes.

### D.2  How to use the call graph module

We use *example_pkg* package as an example and below is the folder structure of it.

```
1    —example_pkg
2        —main.py
3        —sub_folder1
4            —module1.py
5            —module2.py
6        —sub_folder2
```

To construct the call graph of a python application, import and use *CallGraphGenerator* from *scalpel.call_graph.pycg*.

```
1    from scalpel.call_graph.pycg import CallGraphGenerator
2    cg_generator = CallGraphGenerator(["main.py"], "example_pkg")
3    cg_generator.analyze()
```

*CallGraphGenerator* takes two parameters, *entry_points* and *package*. *package* is the root folder of the package that we are generating a call graph for. And *entry_points* is a list of entry point files of the call graph. Now the call graph is generated, many useful functions can be utilized to analyze the package. To output all function calls, *output_edges* can be used:

```
1    edges = cg_generator.output_edges()
```

And one can use *output_internal_mods* and *output_external_mods* to get all internal/external modules.

```
1    internal_mods = cg_generator.output_internal_mods
2    external_mods = cg_generator.output_external_mods
```

To directly operates on the call graph, one can call *output*.
```
1    cg = cg_generator.output()
```

*scalpel.call_graph.pycg* also provides a tool *formats.Simple* to store the call graph results in the JSON format. For more functions, please refer to `https://pypi.org/project/pycg/`.

The tutorial code can be found at: `https://github.com/SMAT-Lab/Scalpel/blob/master/examples/cg_tutorial.py`

```
1  from scalpel.call_graph.pycg import formats
2  import json
3  formatter = formats.Simple(cg)
4  with open("example_results.json", "w+") as f:
5      f.write(json.dumps(formatter.generate()))
```

### D.3  APIs

The API documentation can be found at `https://smat-lab.github.io/Scalpel/scalpel/call_graph.html`

## E  Type Inference

### E.1  Introduction

Type inference module in Scalpel can identify type information for the usage context. As a dynamically type languages, the type of a Python variable is unknown until runtime (known as duck typing), making it difficult to perform type checking. Though benefiting from the coding flexibility for rapid development, Python programs can miss the opportunity to separate data from behavior and detect bugs and errors at an early stage.

Scalpel provides a module *scalpel.typeinfer* for automatic type inference to facilitate static analysis for Python programs. *scalpel.typeinfer* takes a python file or the root folder of a whole package as input, and will output a dictionary of detailed type information for each variable, including function return values and function parameters.

### E.2  How to use Type Inference

Here is a demo input python program we will be using. The piece of code returns current working directory and is named "type_infer_example.py".

```
1  from os import getcwd
2  def my_function():
3      x = "Current working directory: "
4      return x + getcwd()
```

To infer the types of the variables, we will use *TypeInference* class in *scalpel.typeinfer*.

```
1  from scalpel.typeinfer.typeinfer import TypeInference
2
3  inferer = TypeInference(name='type_infer_example.py', entry_point='type_infer_example.py')
4  inferer.infer_types()
5  inferred = inferer.get_types()
```

The first parameter of *TypeInference* is the desired name for the inference analyzer, and the second one is the path to a python file or the root folder of a python package. After instantiating a *TypeInference* analyzer, invoke *infer_types()* method to start the inferring process. *get_types()* will return a list containing inferred type information of all variables. The output is as follows. As shown in the output, the type of the function return value is also inferred,

```
1  [{'file': 'type_infer_example.py', 'line_number': 4, 'function': 'my_function', 'type': {'str'}},
2   {'file': 'type_infer_example.py', 'line_number': 5, 'variable': 'x', 'function': 'my_function', 'type': 'str'}]
```

which is *str*.

The tutorial code can be found at: `https://github.com/SMAT-Lab/Scalpel/blob/master/examples/type_infer_tutorial.py`

### E.3 APIs

The API documentation can be found at `https://smat-lab.github.io/Scalpel/scalpel/typeinfer.html`